\documentclass[a4paper]{jpconf}

\usepackage{graphics}
\usepackage{graphicx}
\usepackage{bm}
\usepackage{amsmath}
\usepackage{amsfonts}
\usepackage{amssymb}
\usepackage{latexsym}
\usepackage{color}

\definecolor{darkred}{rgb}{.8,0,0}

\definecolor{darkblue}{rgb}{0,0,.8}

\newcommand{\be}{\begin{equation}}\newcommand{\ee}{\end{equation}}
\newcommand{\bea}{\begin{eqnarray}}\newcommand{\eea}{\end{eqnarray}}
\newcommand{\brr}{\begin{array}}\newcommand{\err}{\end{array}}
\newcommand{\bit}{\begin{itemize}}\newcommand{\eit}{\end{itemize}}
\newcommand{\ben}{\begin{enumerate}}\newcommand{\een}{\end{enumerate}}

\def\non{\nonumber}

\def\1{{_{1}}}\def\2{{_{2}}}

\begin{document}

\title{Neutrino nature, total and geometric phase}

\author{Antonio  Capolupo }
\address{ Dipartimento di Fisica E.R.Caianiello and INFN Gruppo Collegato di Salerno,
  Universit\'a di Salerno, Fisciano (SA) - 84084, Italy}
\author{S.M. Giampaolo}
\address{Division of Theoretical Physics,  Ruder Bo\v{s}kovi\'{c} Institute, Bijen\u{c}ka cesta 54, 10000 Zagreb, Croatia}

\begin{abstract}
We  study the total and the geometric phase  associated with neutrino mixing and we show that the  phases produced by the neutrino oscillations have different values depending on the representation of the mixing matrix and on the neutrino nature. Therefore the phases represent a possible probe    to distinguish between Dirac and Majorana neutrinos.

\end{abstract}

\section{Introduction}

The phenomenon of neutrino  mixing and oscillation, that has been proved experimentally \cite{Nakamura1}-\cite{Nakamura6}, implies that the neutrino has a mass.
Then the neutrino, being a neutral particle, can be a Majorana particle (a fermion that is its own antiparticle), or a Dirac particle (a fermion different from its antiparticle).
At the moment the neutrino nature  is not established.

A Majorana field is characterized by the presence of the Majorana  phases $\phi_i$ in the mixing matrix which violate  the CP symmetry. These phases cannot be eliminated since the Lagrangian of Majorana neutrinos
    is not invariant under $U(1)$ global transformation.
 By contrast, for Dirac neutrino, the Lagrangian is invariant under $U(1)$ global transformation  and the $\phi_i$
phases can be removed.
  The mixing matrices for Majorana $U_M$ and for Dirac neutrinos $U_D$ can be related  for example by the equation,
$
U_M = U_D \cdot diag(1, e^{i \phi_1}, e^{i \phi_2},..., e^{i \phi_{n-1}} )\,,
$
where  $i = 1,...,n-1 $.
 Other representations of  $U_M$ can be  obtained by the rephasing the lepton charge fields in the charged current weak-interaction Lagrangian \cite{Giunti}. For example, in two flavor neutrino mixing case,  one can consider the following  mixing matrices for Majorana neutrinos
  \bea\label{U1}
 U_{1} =
   \left(
            \begin{array}{cc}
             \cos  \theta   &    \sin  \theta\,  e^{i \phi}\\
             -  \sin  \theta  &  \cos  \theta\,  e^{i \phi}\\
            \end{array}
          \right)
           \,,
           \qquad
          or
         \qquad
U_{2} =    \left(
            \begin{array}{cc}
             \cos  \theta   &    \sin  \theta\,  e^{-i \phi}\\
             -  \sin  \theta\,  e^{i \phi} &  \cos  \theta \\
            \end{array}
          \right)
             \,,
          \eea
where $\theta$ is the mixing angle,   and $\phi$ is the Majorana phase.
It should be noted that, neglecting the dissipation  \cite{CapolupoDiss}, the Majorana phases   do  not affect the neutrino oscillation formulae, being such formulae equivalent for Majorana and for Dirac neutrinos \cite{Pontecorvo}.
Therefore, the oscillation formulae are not useful in the study of the neutrino nature.

Recently, the study of the geometric phase has attracted also a great attention. The  geometric phase
  appears  in the evolution of any quantum state
describing a system characterized by a Hamiltonian defined on a parameter space \cite{Berry:1984jv}--\cite{Pechal}. This phase
 arises in  many physical systems \cite{grafene}--\cite{Blasone:2009xk}   and it has been observed experimentally.

In this paper, we report the results of the study on the total and geometric phases of neutrino presented in Ref.\cite{CapolupoNeutFasi} and we show that, unlike the oscillation formulae, the total phase (and  the dynamical one), generated by the transition between different flavors,   depends on the choice of the matrix $U$. Indeed, different choices of $U$ lead   to different values of the total  phases.
In particular, considering the two flavor neutrino mixing case,   we show that  the use of  the matrix   $\textit{U}_{ 2}$ in Eq.(\ref{U1}) (and of that corresponding to oscillations in a medium), generates values of the phases which are different for Majorana and for Dirac neutrinos.
By contrast, if we consider the $\textit{U}_{  1}$ matrix, all the phases are independent from $\phi$ and  Majorana neutrinos  cannot be distinguished from  Dirac neutrinos.

The paper is organized as follows.
In Section 2 we analyze the total and the geometric phase  for neutrinos   by using different mixing matrces. In Section 3 we report a numerical analysis on the neutrino phases and in Section 4 we give our conclusions.

\section{ Total and geometric phases for neutrinos }

We analyze the neutrinos propagation in   vacuum and  through a  medium.
The matter effects, are taken into account by replacing in the flavor states in vacuum, $\Delta m^{2}$ with $\Delta m_{m}^{2} = \Delta m^{2} R_{\pm}$, and $\sin 2\theta $ with $\sin 2\theta_{m} =  \sin 2\theta /R_{\pm}$.
The coefficients $R_{\pm}$  are, $R_{\pm} = \sqrt{\left(\cos 2\theta  \pm  \frac{2 \sqrt{2 } G_{F} n_{e} E}{\Delta m^{2}}\right)^{2}+ \sin^{2} 2\theta}\,, $  with $+$ for oscillation of antineutrinos and $-$ for oscillations of neutrinos \cite{MSW1,MSW2}.
 In the following, we consider the  flavor states $|\nu_{e}(z) \rangle $ and $| \nu_{\mu}(z) \rangle$ at the $z$   distance given by the mixing matrix $U_2$, with $\theta$ replaced by $\theta_{m}$,
\bea\label{stato1}\non
|\nu_{e}(z) \rangle \!\!& = &\!\! \cos  \theta_{m} e^{ i \frac{\Delta m_{m}^{2} }{4 E} z}  |\nu_{1} \rangle
+
e^{- i \phi} \sin \theta_{m} e^{ -i \frac{\Delta m_{m}^{2} }{4 E} z}|\nu_{2} \rangle ,
\\\label{stato2}
|\nu_{\mu}(z) \rangle \!\!& = &\!\! - e^{ i \phi} \sin \theta_{m} e^{ i \frac{\Delta m_{m}^{2} }{4 E} z}  |\nu_{1} \rangle
+
 \cos \theta_{m} e^{ -i \frac{\Delta m_{m}^{2} }{4 E} z}|\nu_{2} \rangle .
\eea
and we derive the total and the non--cyclic geometric phase \cite{Mukunda}.
For a quantum system whose state vector is $|\psi(s)\rangle$, the geometric phase is defined
as the difference between the total phase $\Phi^{tot}_{\psi} = \arg \langle \psi(s_{1})| \psi(s_{2} )\rangle$ and the dynamic phase
$\Phi^{dyn}_{\psi} = \Im\int_{s_{1}}^{s_{2}}\langle \psi(s)|\dot{\psi}(s)\rangle d s$, i.e.
$
\Phi^{g }   =   \Phi^{tot}_{\psi}  - \Phi^{dyn}_{\psi}.
 $
Here, $s$ is a real parameter such that $s \in [s_1, s_2]$, and
the dot denotes the derivative with respect to  $s$.
For electron neutrino, the geometric phase is
\bea\non\label{fase1}
\Phi^{g }_{\nu_{e}}(z)
 & = &   \arg \left[\langle \nu_{e}(0)| \nu_{e}(z)\rangle  \right] - \Im \int_{0}^{z}  \langle  \nu_{e}(z^{\prime})| \dot{\nu}_{e}(z^{\prime})\rangle  d z^{\prime}
 \\  & = & \arg \left[ \cos \left(\frac{\Delta m_{m}^{2} z}{4 E}\right) + i \cos 2\theta_{m} \sin \left(\frac{\Delta m_{m}^{2} z}{4 E}\right) \right]
- \frac{\Delta m_{m}^{2} z}{4 E} \,\cos 2\theta_{m}  \,.
\eea

 For muon neutrino we have $\Phi^{g }_{\nu_{\mu}}(z) = - \Phi^{g }_{\nu_{e}}(z)$. Eq.(\ref{fase1}) holds both for Majorana and for Dirac neutrinos, indeed it does not depend on the $CP$ violating phase $\phi$ and thus it is independent on the choice of the mixing matrix.
However, we can also consider the following phases due to the neutrino transitions between different flavors,
 \bea  \label{fasemix1}
\Phi_{\nu_{e}\rightarrow \nu_{\mu}}(z)
\!\!& = &\!\!
\arg \left[\langle \nu_{e}(0)| \nu_{\mu}(z)\rangle  \right] - \Im \int_{0}^{z}  \langle  \nu_{e}(z^{\prime})| \dot{\nu}_{\mu}(z^{\prime})\rangle  d z^{\prime}\, ,
\\  \label{fasemix2}
\Phi_{\nu_{\mu}\rightarrow \nu_{e}}(z)
\!\!& = &\!\!
\arg \left[\langle \nu_{\mu}(0)| \nu_{e}(z)\rangle  \right] - \Im \int_{0}^{z}  \langle  \nu_{\mu}(z^{\prime})| \dot{\nu}_{e}(z^{\prime})\rangle  d z^{\prime}\, .
\eea
Eqs.(\ref{fasemix1}) and (\ref{fasemix2}) represent the differences between the total and the dynamic phases generated by the transitions $\nu_{e}\rightarrow \nu_{\mu}$
and $\nu_{\mu}\rightarrow \nu_{e}$, respectively.
 By using the Majorana neutrino states in  Eqs.(\ref{stato2}), we have
 \bea
\label{fasemix1a}
\Phi _{\nu_{e}\rightarrow \nu_{\mu}}(z) &  = &  \frac{3\pi}{2} + \phi
+  \left(\frac{\Delta m_{m}^{2}}{4 E}\,\sin 2\theta_{m}\; \cos \phi\; \right) z\, ,
\\
 \label{fasemix2a}
\Phi _{\nu_{\mu}\rightarrow \nu_{e}}(z)
 &  = &   \frac{3\pi}{2} - \phi
+  \left(\frac{\Delta m_{m}^{2}}{4 E} \sin 2\theta_{m}\; \cos \phi\ \right) z\, .
\eea
 Then, $\Phi _{\nu_{e}\rightarrow \nu_{\mu}} \neq \Phi _{\nu_{\mu}\rightarrow \nu_{e}} $. Although both the total and the dynamic phases depend on $\phi$, the asymmetry between  the transitions  ${\nu_{e}\rightarrow \nu_{\mu}} $ and $  {\nu_{\mu}\rightarrow \nu_{e}}$ is due to  the total phases. Indeed, we have $\Phi^{tot }_{\nu_{e}\rightarrow \nu_{\mu}} =  \frac{3\pi}{2} + \phi$ and $ \Phi^{tot}_{\nu_{\mu}\rightarrow \nu_{e}} = \frac{3\pi}{2} - \phi $ (whereas $\Phi^{dyn }_{\nu_{e}\rightarrow \nu_{\mu}} =  \Phi^{dyn}_{\nu_{\mu}\rightarrow \nu_{e}} =
 \left(\frac{\Delta m_{m}^{2}}{4 E} \sin 2\theta_{m}\; \cos \phi\ \right) z $).
 By contrast, for Dirac neutrinos we have
  \bea
 \label{fasemixD}
\Phi _{\nu_{e}\rightarrow \nu_{\mu}}(z) = \Phi _{\nu_{\mu}\rightarrow \nu_{e}}(z)
=
\frac{3\pi}{2}
+  \left(\frac{\Delta m_{m}^{2}}{4 E}\,\sin 2\theta_{m}\;  \right) z \,,
\eea
and  the total phases reduce to $\Phi^{tot }_{\nu_{e}\rightarrow \nu_{\mu}}(z) = \Phi^{tot}_{\nu_{\mu}\rightarrow \nu_{e}}(z)
= \frac{3\pi}{2} $.
The phases defined in Eqs.(\ref{fasemix1}) and (\ref{fasemix2}) and the total phases depend on the choice of the mixing matrix.
Indeed, if we consider the mixing matrix obtained by $U_{1}$ by replacing $\theta $ with $\theta_m$,  the result of Eq.(\ref{fasemixD})  is obtained also for Majorana neutrinos. Similar results are found for oscillation in vacuum.
Therefore  the  phases $\Phi _{\nu_{e}\rightarrow \nu_{\mu}}$ and $\Phi _{\nu_{\mu}\rightarrow \nu_{e}}$
and the total phases $\Phi^{tot}_{\nu_{e}\rightarrow \nu_{\mu}}$ and $\Phi^{tot }_{\nu_{\mu}\rightarrow \nu_{e}}$
discriminate between the two matrices $U_1$ and $U_2$.

\section{ Numerical analysis.}

In order to   connect  of our results with experiments, we plot in Figs.1 and 2 the total, the geometric  phases and the phases defined in Eqs.(\ref{fasemix1}) and (\ref{fasemix2}) by using the characteristic  values of  experiments such as RENO \cite{Nakamura2} and T2K \cite{Nakamura4}.
\begin{figure}[t]\centering
\begin{picture}(300,180)(0,0)
\put(10,20){\resizebox{8.0 cm}{!}{\includegraphics{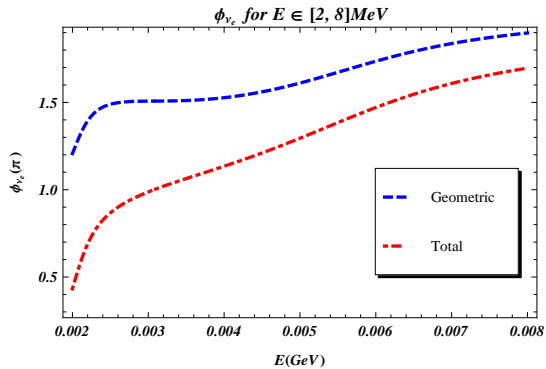}}}
\end{picture}\vspace{-1cm}
\caption{\em (Color online) Plots of the total (the red dot dashed line) and the geometric phases (the blue  dashed line)  of  $ \nu_{e}  $, as a function of the neutrino energy $E$, for a distance length $z = 100 km$.}
\label{pdf}
\end{figure}
In Fig.1 we report  the total and geometric phases associated with the evolution of $\nu_{e}$. We consider the neutrino propagation through the matter and   the values of the parameters of  RENO experiment \cite{Nakamura2}:  neutrino energy   $E \in [2 - 8] MeV$,  electron earth density $n_{e } =10^{24} cm^{-3} $, $\Delta m^{2} = 7.6 \times 10^{-3}eV^{2}$ and   distance $z = 100 km$.
\begin{figure}[t]\centering
\begin{picture}(300,180)(0,0)
\put(10,20){\resizebox{8.0 cm}{!}{\includegraphics{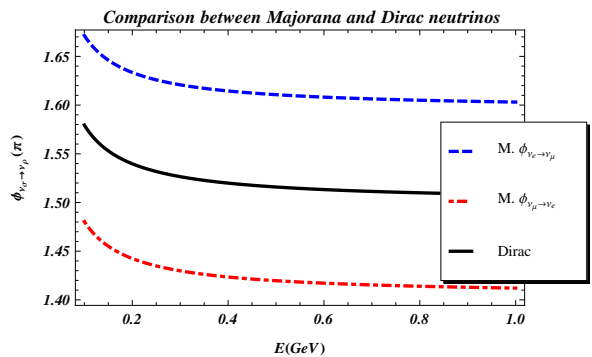}}}
\end{picture}\vspace{-1cm}
\caption{\em (Color online) Plot of the  phases $\Phi _{\nu_{e}\rightarrow \nu_{\mu}} $ (the blue dashed line)  and $\Phi _{\nu_{\mu}\rightarrow \nu_{e}} $ (the red dot dashed line) for Majorana neutrinos as a function of the neutrino energy $E$, for a distance length $z = 300 km$. The   phases $\Phi _{\nu_{e}\rightarrow \nu_{\mu}} =\Phi _{\nu_{\mu}\rightarrow \nu_{e}} $ for Dirac neutrinos  is represented by the black solid line.}
\label{pdf}
\end{figure}
In Fig.2 we report  the   phases $\Phi _{\nu_{e}\rightarrow \nu_{\mu}} $ and $\Phi _{\nu_{\mu}\rightarrow \nu_{e}} $, by assuming $E \sim 1 GeV$ and   $z = 300 km$, which are values compatible with the parameters of $T2K$ experiment \cite{Nakamura4}. Moreover we consider $\phi = 0.3$, and the values of $n_e$ and $\Delta m^{2}$ considered above.

\section{ Conclusions.}

We analyzed  the total and the geometric phases generated in the evolution  of the neutrino.
We have shown that  for Majorana neutrinos  the   phases  due to a transition between different neutrino flavors
  take different values depending on the representation of the mixing matrix and on the nature of neutrinos.
 By considering the mixing matrix $U_2$, we obtained  for Majorana neutrinos, $\Phi_{\nu_{e}\rightarrow \nu_{\mu}} \neq \Phi_{\nu_{\mu}\rightarrow \nu_{e}} $
(and   $\Phi^{tot}_{\nu_{e}\rightarrow \nu_{\mu}} \neq \Phi^{tot}_{\nu_{\mu}\rightarrow \nu_{e}} $), that reveals an asymmetry in the transitions
$\nu_{e}\rightarrow \nu_{\mu}$ and $\nu_{\mu}\rightarrow \nu_{e}$. This asymmetry disappears for Dirac neutrinos.
On the contrary, by using $U_1$,
 we have $\Phi_{\nu_{e}\rightarrow \nu_{\mu}} = \Phi_{\nu_{\mu}\rightarrow \nu_{e}} $, (and $\Phi^{tot}_{\nu_{e}\rightarrow \nu_{\mu}} = \Phi^{tot}_{\nu_{\mu}\rightarrow \nu_{e}} $) both for Dirac and Majorana neutrinos and
 nothing can be said on the neutrino natures.
 We presented a numerical analysis   by using the characteristic parameters of RENO and T2K experiments and we have obtained values for the neutrino phases which, in principle, are detectable. Our results pave  the way for a completely new method to study the nature of neutrinos.
  In our discussion, the quantum field theory effects on particle mixing \cite{Blasone:1998hf}–-\cite{Capolupo:2010ek},
 can be safely neglected  \cite{CapolupoPLB2004}.

\section*{Acknowledgements}
A.C. acknowledges partial financial support from MIUR and INFN and  the COST Action CA1511 Cosmology
and Astrophysics Network for Theoretical Advances and Training Actions (CANTATA)
supported by COST (European Cooperation in Science and Technology).
S.M.G. acknowledge support by the H2020 CSA Twinning project No. 692194, "RBI-T-WINNING''.


\section*{References}



\begin{thebibliography}{999}




\bibitem{Nakamura1}
 An F P  et al. [Daya-Bay Collaboration]  2012
{\it  Phys. Rev. Lett.} {\bf 108}, 171803

\bibitem{Nakamura2}
 Ahn J K  et al. [RENO Collaboration]  2012
Experiment,” {\it Phys. Rev. Lett.}{\bf 108}, 191802

\bibitem{Nakamura3}
 Abe  Y  et al. [Double Chooz Collaboration]  2012
{\it Phys. Rev. Lett.} {\bf 108}, 131801

\bibitem{Nakamura4}
 Abe K  et al. [T2K Collaboration]  2011
{\it Phys. Rev. Lett.} {\bf 107}, 041801

\bibitem{Nakamura5}
 Adamson P  et al. [MINOS Collaboration]  2011
{\it Phys. Rev. Lett.} {\bf 107}, 181802


\bibitem{Nakamura6}
 Nakamura K and Petcov S T  2012
{\it Phys. Rev. D} {\bf 86}, 010001


\bibitem{Giunti}
 Giunti C  2010 {\it Phys. Lett. B} {\bf 686}, 41


\bibitem{CapolupoDiss}
Capolupo A, Giampaolo S M and Lambiase G 2018
  arXiv:1807.07823 [hep-ph]


\bibitem{Pontecorvo}
Bilenky S M and Pontecorvo  B  1978 Phys. Rep. {\bf 41}, 225

 

\bibitem{Berry:1984jv}
 Berry M V  1984
 {\it Proc.\ Roy.\ Soc.\ Lond. A} {\bf 392}, 45
%
 \bibitem{Berry:1984jv1}
 Aharonov Y and Anandan J  1987
  {\it Phys.\ Rev.\ Lett.} {\bf 58}, 1593
%
%
 \bibitem{Berry:1984jv2}
Samuel J and Bhandari R  1988  {\it Phys.\ Rev.\ Lett.} {\bf 60}, 2339

%
 \bibitem{Berry:1984jv4} 
 Pancharatnam S  1956
 Proc.\ Indian Acad.\ Sci.\ A {\bf 44}, 1225

\bibitem{Shapere}
Shapere A and Wilczek F  1989  Geometric Phases in Physics, {\it  World Scientific}, Singapore.
%

%
\bibitem{Shapere2} Garrison J C and Wright E M  1988
  {\it Phys. Lett. A} {\bf 128}, 177
%

\bibitem{Shapere4} 
Pati A K  1995
 {\it J. Phys. A} {\bf 28}, 2087
%
%
\bibitem{Shapere5}
  Pati A K  1995
 {\it Phys. Rev. A} {\bf 52}, 2576

\bibitem{Mukunda} Mukunda N and Simon R  1993  {\it Ann. Phys.}(N.Y) {\bf 228}, 205

\bibitem{Mostafazadeh}
  Mostafazadeh A  1999
 {\it J. Phys. A} {\bf 32}, 8157
%
\bibitem{Mostafazadeh1} Anandan J  1988
 {\it Phys. Lett. A} {\bf 133}, 171

\bibitem{Tomita} Tomita A and Chiao R Y  1986  {\it Phys.\ Rev.\ Lett.} {\bf 57}, 937

\bibitem{Jones}
Jones J A, Vedral V, Ekert A  and Castagnoli G  2000 {\it Nature} {\bf 403}, 869

\bibitem{Leek}
Leek P J  et. al  2007  {\it Science} {\bf 318}, 1889

\bibitem{Leek1} Neeley M  et al.  2009  {\it Science} {\bf  325}, 722

\bibitem{Pechal}
 Pechal M  et al.  2012  {\it Phys. Rev. Lett.} {\bf  108}, 170401

\bibitem{grafene}
Zhang Y, Tan Y W, Stormer H L and Kim P  2005
 {\it Nature} {\bf  438}, 201-204

 \bibitem{Capolupo:2013xza1}
 Falci G  et al.  2000
{\it Nature} {\bf 407}, 355-358

\bibitem{Capolupo:2013xza2}
 Mottonen M, Vartiainen J J and Pekola J P  2008
 {\it Phys. Rev. Lett.} {\bf  100}, 177201

 \bibitem{Capolupo:2013xza3}
 Murakawa H et al.  2013
 {\it Science}, {\bf  342}, Issue 6165,  1490-1493


  \bibitem{Capolupo:2013xza12}
 Xiao D et al.  2010 {\it Rev. Mod. Phys.} {\bf 82}, 1959

\bibitem{Capolupo:2013xza13}
Capolupo A and Vitiello G  2013
 {\it Adv.\ High Energy Phys.} {\bf 2013}, 850395
 %
\bibitem{Capolupo:2013xza14}
 Capolupo A and Vitiello G  2013
 {\it Phys.\ Rev.\ D} {\bf 88}, 024027
  %
 \bibitem{Capolupo:2013xza15}
 Capolupo A and Vitiello G 2015
{\it  Adv.\ High Energy Phys.}
{\bf 2015}, 878043
%
\bibitem{Capolupo:2013xza16}
   Bruno A, Capolupo A, Kak S, Raimondo G and Vitiello G  2011
  {\it Mod.\ Phys.\ Lett.\ B} {\bf 25}, 1661
%
\bibitem{Capolupo:2013xza17}
Hu J and Yu J  2012
 {\it Phys.\ Rev.\  A} {\bf 85}, 032105




 \bibitem{Blasone:2009xk1}
 Blasone M, Capolupo A, Celeghini E and Vitiello G  2009
  {\it Phys.\ Lett.\ B} {\bf 674}, 73

\bibitem{Blasone:2009xk2}
Joshi S and Jain S  R  2016 {\it Phys. Lett. B} {\bf 754}, 135

 \bibitem{Blasone:2009xk3}
  Johns L  and Fuller G M 2017 {\it Phys. Rev. D} {\bf 95}, 043003

 \bibitem{Blasone:2009xk4}
Capolupo A, Lambiase G and Vitiello G 2015
  {\it Adv.\ High Energy Phys.\ } {\bf 2015}, 826051



\bibitem{Beatrix}
Bertlmann R A, Durstberger K, Hasegawa Y and Hiesmayr B C  2004
{\it Phys.Rev.A} {\bf 69}, 032112


\bibitem{Blasone:2009xk}
  Capolupo A 2011
 {\it Phys.\ Rev.\ D} {\bf 84}, 116002


\bibitem{CapolupoNeutFasi}
Capolupo A, Giampaolo S M, Hiesmayr B C and Vitiello G  2018
  {\it Phys.\ Lett.\ B} {\bf 780}, 216





\bibitem{MSW1}
Mikheev S P and Smirnov A Yu  1985
  {\it Sov. J. Nuc. Phys.}  {\bf 42} (6): 913–917

\bibitem{MSW2}
Wolfenstein L  1978
  {\it Phys. Rev.  D} {\bf 17} (9): 2369


\bibitem{Blasone:1998hf}
Blasone M, Capolupo A and Vitiello G 2002
 {\it  Phys.\ Rev.\ D } {\bf 66}, 025033  and references therein
%

\bibitem{Blasone:1998hf1}
 Blasone M, Capolupo A,   Romei O  and  Vitiello G 2001
 {\it Phys.\ Rev.\ D} {\bf 63}, 125015

\bibitem{CapolupoPLB2004}
 Capolupo A, Ji C-R,  Mishchenko Y  and  Vitiello G 2004
 {\it Phys.\ Lett.\ B } {\bf 594}, 135

\bibitem{Blasone:2006jx}
 Blasone M, Capolupo A, Terranova F and Vitiello G 2005
  {\it  Phys.\ Rev.\ D}  {\bf 72}, 013003

 \bibitem{Blasone:2006jx1}
Blasone M, Capolupo A, Ji C-R and Vitiello G 2010
 {\it  Int.\ J.\ Mod.\ Phys.\ A } {\bf 25}, 4179


\bibitem{Capolupo:2006et}
  Capolupo A 2018
  {\it Adv.\ High Energy Phys.   } {\bf 2018}, 9840351


\bibitem{Capolupo:2006et0}
Capolupo A 2016
 {\it Adv. High Energy Phys. } {\bf 2016}, 8089142

  \bibitem{Capolupo:2006et1}
  Capolupo A, Capozziello S and Vitiello G 2009
 {\it   Phys.\ Lett.\ A } {\bf 373}, 601
%
\bibitem{Capolupo:2006et2}
  Capolupo A, Capozziello S and Vitiello G 2007
 {\it  Phys.\ Lett.\ A}  {\bf 363}, 53
%
 \bibitem{Capolupo:2006et3}
  Capolupo A, Capozziello S and Vitiello G 2008
   {\it Int.\ J.\ Mod.\ Phys.\ A}  {\bf 23}, 4979
  %
\bibitem{Capolupo:2006et4}
Blasone M, Capolupo A, Capozziello S and  Vitiello G 2008
 {\it Nucl.\ Instrum.\ Meth.\  A}  {\bf588}, 272
%

\bibitem{Blasone:2009vk}
  Blasone M,  Capolupo A and  Vitiello G  2010
  {\it Prog.\ Part.\ Nucl.\ Phys.\  } {\bf 64}, 451

\bibitem{Blasone:2004yh}
   Blasone M,  Capolupo A,  Capozziello S,  Carloni S and  Vitiello G  2004
   {\it Phys.\ Lett.\ A }  {\bf 323}, 182


\bibitem{CapolupoAssioni}
Capolupo A, De Martino I, Lambiase G and  Stabile A  2019 {\it Axion--photon mixing in quantum field theory and vacuum energy},
 {\it Phys. Lett. B}, in press, https://doi.org/10.1016/j.physletb.2019.01.056.


%
\bibitem{Capolupo:2013af}
  Capolupo A and  Di Mauro M 2013
{\it Acta Phys.\ Polon.\ B} {\bf 44}, 81

\bibitem{Capolupo:2010ek}
Capolupo A, Di Mauro M and  Iorio A 2011
 {\it  Phys.\ Lett.\ A } {\bf 375}, 3415



\end{thebibliography}
\end{document}